\documentclass[aps,prc,twocolumn,tightenlines,showpacs,
superscriptaddress,nofootinbib,14pt]{revtex4}
\usepackage[figuresright]{rotating}

\begin{document}

\title{ICP-SFMS search for long-lived naturally-occurring heavy, superheavy and superactinide nuclei compared to AMS experiments}

\author{A. Marinov}
\email{marinov@vms.huji.ac.il} \altaffiliation{Fax:
+972-2-6586347.}
 \affiliation{Racah Institute of Physics, The Hebrew
University of Jerusalem, Jerusalem 91904, Israel}
\author{A. Pape}
\affiliation{IPHC-UMR7178, IN2P3-CNRS/ULP, BP 28, F-67037
Strasbourg cedex 2, France}
\author{Y. Kashiv}
\affiliation{Racah Institute of Physics, The Hebrew University of
Jerusalem, Jerusalem 91904, Israel}
\author{D. Kolb}
\affiliation{Department of Physics, University GH Kassel, 34109
Kassel, Germany}
\author{L. Halicz}
\affiliation{Geological Survey of Israel, 30 Malkhei Israel St.,
Jerusalem 95501, Israel}
\author{I. Segal}
\affiliation{Geological Survey of Israel, 30 Malkhei Israel St.,
Jerusalem 95501, Israel}
\author{R. Brandt}
\affiliation{Kernchemie, Philipps University, 35041 Marburg,
Germany}

\date{September 22, 2011}


\begin{abstract}
Negative results obtained in AMS searches by Dellinger et al. on mostly unrefined ores have led them to conclude that the very heavy long-lived species found in chemically processed samples with ICP-SFMS by Marinov et al. are artifacts.  We argue that it may not be surprising that results obtained from small random samplings of inhomogeneous natural minerals would contrast with concentrations found in homogeneous materials extracted from large quantities of ore.  We also point out that it is possible that the groups of counts at masses 296 and 294 seen by Dellinger et al. could be, within experimental uncertainties, due to $^{296}$Rg and $^{294}$eka-Bi in long-lived isomeric states.  In such case, the experiments of Dellinger et al. lend support to the experiments of Marinov et al.

\end{abstract}

\maketitle

In a recent paper entitled  "Ultrasensitive search for long-lived superheavy nuclides in the mass range A = 288 to A = 300 in
natural Pt, Pb, and Bi", Dellinger et al. \cite{del11} searched for superheavy elements, mainly in native samples  of
Pt, Pb, and Bi, using the AMS technique. Negative results were claimed in this and in their previous investigations on Th \cite{del10} and Au \cite{del11a}, with  upper limits of relative abundances in the range of 5x10$^{-13}$ to 10$^{-16}$ g/g. The authors then  concluded that the observation of long-lived isomeric states, by measuring accurate masses using ICP-SFMS, in various heavy, superheavy and superactinide nuclei are artifacts. Such isomeric states have been reported in the neutron-deficient  $^{211,213,217,218}$Th isotopes (abundance (1-10)x10$^{-11}$ relative to $^{232}$Th) \cite{mar07}, in the  superheavy nuclei $^{261,265}$Rg (abundance of (1-10)x10$^{-10}$ relative to Au) \cite{mar09,mar11}\footnote{In \cite{mar11} an enrichment of Rg relative to Au of three to four orders of magnitude has been achieved.}, and in the  superactinide nucleus $^{292}$eka-Th (abundance of (1-10)x10$^{-12}$ relative to  $^{232}$Th) \cite{mar08}.

An important difference between our works \cite{mar07,mar09,mar11,mar08} and those of Dellinger et al. \cite{del11,del10,del11a} is that  we  used processed  Au and Th starting materials, and Dellinger et al., except for a few cases, used raw minerals. It may not be surprising that results obtained with random samplings of a few mg would contrast with concentrations found in processed homogeneous starting materials which probe a large amount of natural ore. This factor could be more important than presumably very little differences in separation factors between an eka-element and its lower homologue that might occur during the purification of the lower homologue from the ore.
 The cases where Dellinger et al. studied  processed samples were ThO$_{2}$ \cite{del10}, Pt, PbS (galena) and Bi \cite{del11}. The latter three are irrelevant for comparison with our results since we  did not measure these elements. (However
 see below.)       As for  ThO$_{2}$, we do not think, that based on a single measurement
 using a very complicated system, one can conclude that all our measurements \cite{mar07,mar09,mar11,mar08} done
 with the relatively straightforward ICP-SFMS system are artifacts.
It would be more convincing to point out a weakness in our measurements, which neither we nor Dellinger et al. \cite{del11,del10,del11a}  have been able to find.

  It is claimed \cite{del10} that they checked the efficiency of their AMS system by measuring the ratio of $^{228}$Th/$^{232}$Th. For $^{232}$Th in equilibrium with its daughters this ratio should be equal to the ratio of the corresponding half-lives which is 1.4x10$^{-10}$. However, when one measures the mass 228 with AMS (or with ICP-SFMS) one measures, together with $^{228}$Th, also $^{228}$Ra, which belongs to the same radioactive chain. Its half-life is 3.0 times longer than that of $^{228}$Th. In addition, it is possible that the formation  of negative ions of RaO$_{2}$ is higher, perhaps much higher, than for ThO$_{2}$. According to the "Negative-Ion Cookbook" of Middleton \cite{mid89},  the production of negative  ThO$_{2}$ ions is quite poor, and the  maximum current  measured by him was 50 nA.  It is not clear how Dellinger et al. obtained an average  current of about 320 nA  of negative ThO$_{2}$ ions (table 2 of \cite{del10}).

Another comment we would  like to make is related to the conclusion of Dellinger et al. that based on their results,
  there are no naturally-occurring SHEs.
 In addition to what  was mentioned above \cite{mar07,mar09,mar11,mar08} and also in \cite{mar03}, it seems to us that even their results could indicate the contrary.
   We refer in particular to Fig. 4(b) in \cite{del11a} on $^{296}$Rg  and Fig. 11(b) in \cite{del11} on $^{294}$eka-Bi. Both spectra are clean, without pile-up. In the first one  there is a group  of five events and in the second one there is a group of six events, both very close to the estimated positions of $^{296}$Rg and $^{294}$eka-Bi, respectively. These groups were ignored by the authors on the basis of their measured residual energies. In the first case the peak appears at a residual energy of about 10.5 MeV, where according to the authors, its center should appear at 12.0 MeV.  In the second case of $^{294}$eka-Bi this peak appears at 11.8 MeV, where its calculated position should be at 13.0 MeV.  Such differences  of 1.5 and 1.2 MeV out of predicted energy loss
   in the detector window of 12 and 10.5 MeV (about half of the initial energies of the ions of 24.0 and 23.5 MeV, respectively) could be due
   to experimental and theoretical uncertainties.  Besides in  window thickness and energy
   calibration of the AMS detector, there could be uncertainties in the energy loss and range when extrapolated to unstudied heavy species like Rg and eka-Bi.  In addition, the appreciable scatter of these ions in the detector window, due to their large energy loss,
 decreases the residual energy of the ions in the detector.  In conclusion, a residual energy of 10.5 MeV instead of 12 MeV in the case of  $^{296}$Rg, and 11.8 MeV instead of 13.0 MeV in the case of $^{294}$eka-Bi, when the total energy loss in
the window is about 12 MeV, could be within the uncertainties inherent in the experiments.

If $^{296}$Rg and $^{294}$eka-Bi have been observed in these experiments, then it is a very important result. With Z = 111 and N = 185 for $^{296}$Rg, Z = 115 and N = 179 for $^{294}$eka-Bi, they are in the center of the  island of stability predicted for nuclei in their normal g.s. Since if found in natural materials, their half-lives should be
 $\geq$ 10$^{8}$ y, or otherwise they would have decayed away. However, the predicted half-lives for $^{296}$Rg and $^{294}$eka-Bi in their normal g.s. are 4.5x10$^{6}$ and 1.0x10$^{4}$ s, respectively \cite{mol97}. A consistent interpretation is that, like in $^{211,213,217,218}$Th \cite{mar07}, $^{261,265}$Rg \cite{mar09,mar11} and $^{292}$eka-Th \cite{mar08}, long-lived isomeric states exist in $^{296}$Rg and $^{294}$eka-Bi.

In summary, we have pointed out that the article of Dellinger et al. \cite{del11} does not show  that the observation of long-lived isomeric states in neutron-deficient Th isotopes \cite{mar07}, in superheavy $^{261,265}$Rg nuclei \cite{mar09,mar11}, and in the superactinide nucleus $^{292}$eka-Th \cite{mar08}, are artifacts. It is also pointed out that long-lived $^{296}$Rg and $^{294}$eka-Bi may have been observed by them.  If so, based on lifetimes, it is argued that these species would not be in their normal ground state, but rather in long-lived isomeric states as have been reported earlier \cite{mar07,mar09,mar11,mar08}. These results may add credibility to the original discovery of long-lived isomeric states in  naturally-occurring heavy, superheavy and superactinide nuclei.

\end{document}